\documentclass[10pt,aps,twocolumn,prl,floatfix,superscriptaddress]{revtex4-1}
\usepackage{graphicx}
\usepackage{color}
\usepackage{amssymb,amsfonts,amsmath,amsthm}
\usepackage{bm}  
\usepackage{epstopdf}

\newcommand{\psizk}{|\psi_0\rangle}
\newcommand{\psizb}{\langle\psi_0|}

\begin{document}

\title{Boundaries and Unphysical Fixed Points in Dynamical Quantum
  Phase Transitions} 

\author{Amina Khatun} 

\affiliation{Institute of Physics, Bhubaneswar 751005, India}
\email{amina.burd\string@gmail.com}

\author{Somendra M.  Bhattacharjee}

\affiliation{Department of Physics, Ashoka University, Sonepat,
  Haryana - 131029, India}
\email{somendra.bhattacharjee\string@ashoka.edu.in}


\newcommand{\hlattice}{%
\begin{figure}[htbp]
\includegraphics[width=0.9\linewidth]{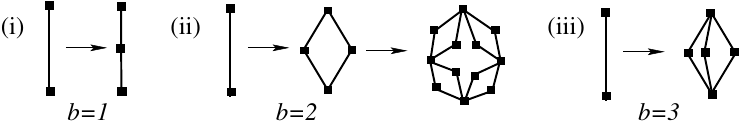}
\caption{Construction of hierarchical lattices.  The sites are
  represented by squares.  Replace each bond by a motif of $b$ branches.
  {\it (i)} $b=1$, {\it (ii)} $b=2$ (diamondlike motif), and {\it
    (iii)} $b=3$.  Three generations are shown for $b=2$.  }
\label{fig:1}
\end{figure}
}%

\newcommand{\bzeros}{%
 \begin{figure}[!htb]
\begin{center}
 \includegraphics[clip]{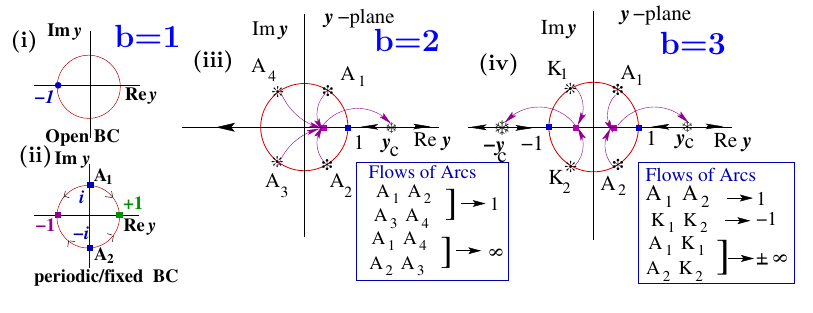}  
\end{center}
\caption{Zeros of $L(y)$ in the complex-$y$ plane, and
  RG flows.  The red circle is the unit circle (UC) for time
  evolution.  For $b=1$, (i) only one zero at $y=-1$ for open BC,
  while {(ii)} the zeros populate the imaginary axis for periodic or
  fixed BC.  {(iii)} For $b=2$, the zeros meet the UC at four points,
  A$_p, p={1,2,3,4}$. Under RG, UC flows to the positive real axis,
  taking each A$_{p}$ to $y_c=3.38298...$.  { (iv)} For $b=3$, the
  four meeting points are of two types; A$_{1}$, A$_{2}$ flow to
  $y_c=2.05817...$, while K$_{1}$, K$_{2}$ to $-y_c<0$. See Fig.
  \ref{fig:22}}
 \label{fig:2}
 \end{figure}
}%

\newcommand{\bjulia}{%
 \begin{figure}[!htb]
\begin{center}
 \includegraphics[clip]{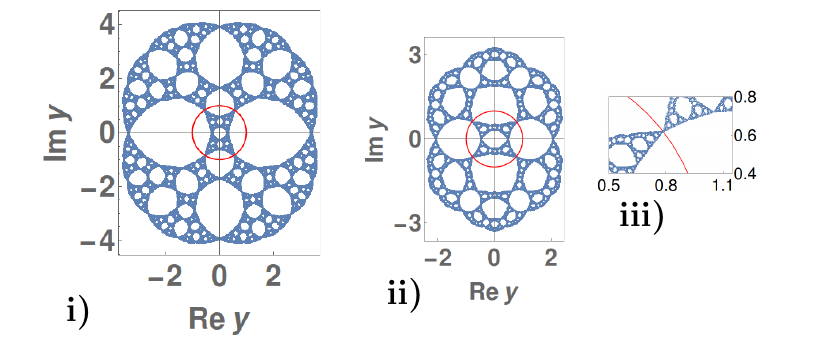}  
\end{center}
\caption{Zeros of $L(y)$ as Julia sets in the
  complex-$y$ plane for (i) $b=2$, and (ii) $b=3$.  See Fig.
  \ref{fig:2}.  The zeros pinch UC at four points.  {(iii)} For $b=3$,
  zoomed view of the region near A$_1$ of Fig. \ref{fig:2}(iv) }
 \label{fig:22}
 \end{figure}
}%
 
\newcommand{\onedchain}{%
\begin{figure}[htbp]

\begin{center}
\includegraphics[scale=1,trim=0 10pt 0 0,clip]{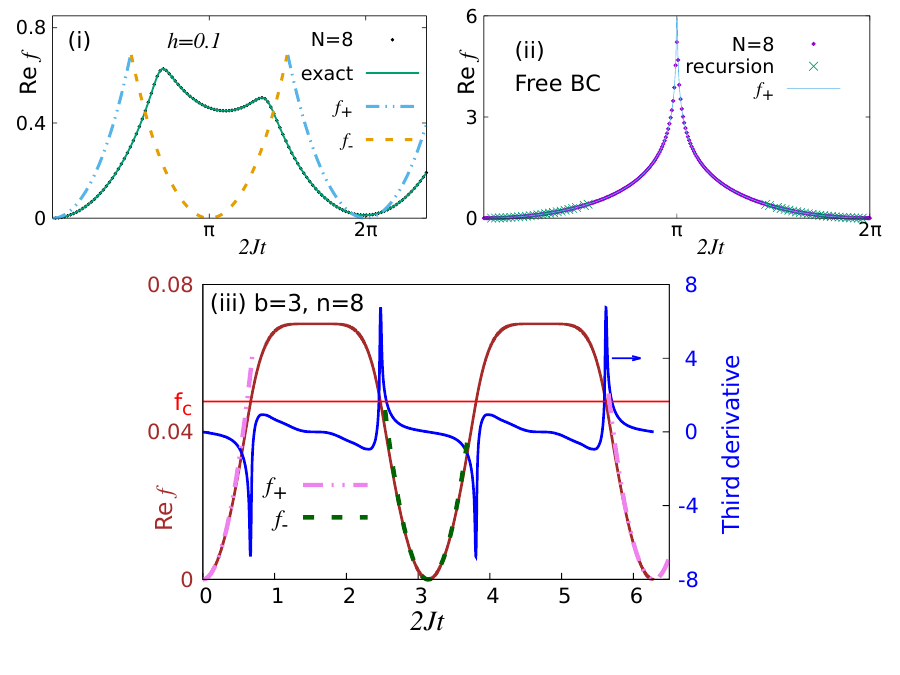}
\end{center}

\caption {Rate function ${\rm Re}\,f(y=e^{i2Jt})$ vs
  $2Jt (=\theta)$.  (i) $b=1$: for a chain of $N=8$ sites with
  boundary field $h=0.1$.  ${\rm Re}\,f(y)$ from a direct time
  evolution {(using {\small{MATLAB}}) } (discs) agrees with
  Eq.\,(\ref{eq:51}) (solid green line).  The $N\to\infty$ limit is
  shown by the blue dash-dot ($f_+$) and orange dashed ($f_-$) lines,
  with DQPT at A$_1$,   $\theta=\pi/2$, and A$_2$,  $\theta=3\pi/2$
  [Fig.~\ref{fig:2}(iv)].  (ii) Same as (i) but without the boundary
  field ($h=0$): time evolution data from {\small MATLAB} agree with
  $f_+$ (solid line) and with {the values} (green crosses) from
  Eq.\;~(\ref{eq:b}) [which fails for $\theta \in (\pi/2,3\pi/2)$].
  (iii) For $b=3$: $f$ (thin brown line) from Eq.\;~(\ref{eq:b}) for
  $n=8$ with ++ spins at the two boundary points.  The intersections
  of the horizontal line at $f_c$ (the critical value of $f$,
  Supplemental Material \cite{suppl}) with $f$ locates the
  transition points A$_{1,2}$ and K$_{1,2}$ [Fig.~\ref{fig:2}(vi)].
  At these points $d^3 ({\rm Re}\,f)/dt^3$ diverges (solid blue line).
  Around $\theta=0$ and $\pi$, $f(y)$ matches with $f_{+}$ (magenta
  dash-dot) and $f_{-}$ (green dashed lines), respectively.  }
\label{fig:4}
\end{figure}
}%

\begin{abstract}
  We show that dynamic quantum phase transitions (DQPT) in many
  situations involve renormalization group (RG) fixed points that are
  unphysical in the context of thermal phase transitions.  In such
  cases, boundary conditions are shown to become relevant to the
  extent of even completely suppressing the bulk transitions.  We
  establish these by performing an exact RG analysis of the quantum
  Ising model on scale-invariant lattices of different dimensions, and
  by analyzing the zeros of the Loschmidt amplitude.  Further
  corroboration of boundaries affecting the bulk transition comes from
  the three-state quantum Potts chain, for which we also show that the
  DQPT corresponds to a pair of period-2 fixed points.
\end{abstract}

\maketitle

Dynamical quantum phase transitions (DQPT), a recent discovery of
phase transitions, often periodic, in large quantum systems during
time evolution \cite{heylprl,heyl2,comm1}, have generated a lot of
interest because here time itself acts as the parameter inducing the
transitions.  Also, to be at a transition point, only time needs to be
chosen properly without any requirement of fine-tuning of system
parameters, unlike thermal transitions \cite{comm11}.  The signature
of DQPT is the nonanalytic behaviour of various quantities in time
around critical times $t_c$'s.  These transitions have now been shown
in many models, like the transverse-field Ising model (TFIM), spin
chains, quantum Potts models, the Kitaev model, and many
others\cite{heylprl,heyl2,Andras,ks17,zvyagin,pozsgay,heylreview}, and
also observed experimentally\cite{expt2,experiment}.  In spite of
being a zero-temperature quantum phenomenon, DQPT is not determined by
the quantum phase transitions of the system but rather seems related
to the classical thermal criticalities of an associated
system\cite{heylreview}.  However, despite the use of many techniques
so far, very few exact results are known on the scaling and
universality in DQPT \cite{heylreview}.  Moreover, the natures of the
possible phases and the transitions remain to be properly classified,
e. g., whether only equilibrium phases and transitions would suffice
or there can be specialities of its own\cite{heylreview}.

A general approach for phase transitions is the renormalization group
(RG) framework \cite{cardy} in terms of length-dependent effective
parameters and their flows to the fixed points (FP), with the stable
FPs determining the allowed phases, and the unstable ones (or
separatrices) the phase transitions.  In this Letter, we adopt an {\it
  exact} RG scheme for TFIM and the three-state quantum Potts chain
(3QPC).  Our exact results establish that there are DQPTs involving
FPs that are {\it unphysical} in traditional thermal transitions.
Second, we show that, for those unphysical FPs, boundary conditions
(BC) are relevant and can even lead to a suppression of the
transitions completely, unlike thermal cases where BCs do not affect
the bulk transitions.  Another surprising result is the emergence of a
pair of period-2 FPs, never seen in the thermal context, that controls
the DQPT in 3QPC, in contrast to the zero-temperature FP \cite{heyl2}
for the Ising DQPT case.  In short, our exact results bring out
several distinctive features of DQPT, not to be found in equilibrium
transitions.

If a quantum system, with Hamiltonian $H$, is prepared in a
noneigenstate $|\psi_0\rangle$ and suddenly allowed to evolve, then
the probability for the system to be in state $|\psi_0\rangle$ after
time $t$ is given by $P(t)=|L(t)|^2\sim e^{-N\lambda(t)},$ where
\begin{equation}
  L(t)=\psizb e^{-itH}\psizk \sim e^{-Nf(t)}, \quad (\hbar=1), \label{eq:9}
\end{equation}
is the Loschmidt amplitude with $f(t)$ and $\lambda(t)=2 {\rm Re}
f(t)$ as the large-deviation rate functions\cite{touchet} for a large
system of $N (\to\infty)$ degrees of freedom.  Often, $\lambda(t)$ and
$f(t)$ show phase-transition-like nonanalyticities at time $t=t_c$.
These phase transitions in time are the
DQPTs\cite{zvyagin,heylreview}.

TFIM is defined on a lattice as $H_{\rm I}=H+H_{\Gamma}$, where
\begin{equation}
  \label{eq:3}
   H=-J\sum_{\langle jk\rangle} \sigma_j^{\rm z}\;\sigma_k^{\rm z},\quad
H_{\Gamma}=-\Gamma \sum_j \sigma_j^{\rm x},
\quad (J,\Gamma>0),
\end{equation}
$\sigma^{\alpha}_j$ being the Pauli matrices ($\alpha=$ x, y, z) at
lattice site $j$, and $\langle jk\rangle$ denoting nearest
neighbours\cite{bkc}.  The interaction favours an aligned state in the
z direction \cite{spectrum}, and $H_{\Gamma}$ is the transverse field
term that aligns the spins in the x direction.  We may add a boundary
term given by $H_{\rm B}=-h(\sigma_1^{\rm z}+\sigma_N^{\rm z}),$ where
the boundary field $h$ acts only on the first and the $N$th spins.
Two special cases are $h=0$ and $h\to\infty$ corresponding to open BC
and fixed BC (both up in the z-direction) respectively.  For periodic
BC in one dimension, $H_B=-J \sigma_1^{\rm z}\;\sigma_N^{\rm z}.$

The TFIM is prepared in a product state $\psizk$ \cite{commx} with all
spins aligned in the x direction, e.g., by $\Gamma\to\infty.$ At time
$t=0$, we set $\Gamma=0$.  So, the magnet evolves with $H$ of
Eq.\,(\ref{eq:3}) and any boundary term mentioned above.  This is the
particular sudden quench we use in this Letter.  By expressing $\psizk$
in terms of the eigenstates of $H$, the Loschmidt amplitude and the
rate function per bond can be expressed as \cite{heylreview,spectrum},
\begin{equation}
   \label{eq:4}
 L(y)= 2^{-N}\sum_C y^{-E_C/2J},
f(y)=-N_B^{-1}~ \ln L(y),
\end{equation}
respectively, where $y=e^{2zJ}, N_B$ is the number of bonds, and, for
generality, $z$ is taken as a complex number.  $L(y)$ is an analytic
continuation of the partition function of the traditional nearest
neighbour Ising model\cite{huang} defined for $1\leq y <\infty$ on the
real positive axis ($z=\beta$ being the inverse temperature).  The
quantum time evolution in Eq.\,(\ref{eq:9}) is given by the unit
circle $|y|=1$ ($y=e^{i2Jt}$) in the complex $y$ plane.  A phase
transition---defined as the point of nonanalyticity of $f$---is
expected along the unit circle if there are zeros or limit points of
zeros of $L(y)$ on the path\cite{huang}.  An isolated zero on the
circle, in contrast, just indicates orthogonality of the evolved and
the initial states.  The $S^1$ (circle) topology guarantees (via
winding numbers) that, if there are zeros on the circle, there will be
periodic transitions in time.

In one-dimensional TFIM and 3QPC, similar DQPT occurs, viz., linear
kinks in $f(t)$, despite the absence of any thermal
transitions\cite{heylprl,ks17}.  For the two-dimensional TFIM, DQPT
was found to be the same as the two dimensional Ising critical
point\cite{heyl2}.  However, the generality of these results has not
yet been established.  In this context, we focus on a class of exactly
solvable models that would help us in alienating the specialities of
DQPT.

We choose scale-invariant lattices for which the real space
renormalization group (RSRG) can be implemented exactly.  The lattices
are constructed hierarchically by replacing a bond iteratively by a
diamondlike motif of $b$ branches\cite{hier,smb} as shown in
Fig.~\ref{fig:1}.  Such lattices appear naturally in approximate RSRG
for usual lattices.  Three cases are considered here, {\it (i)} $b=1$
corresponding to a one-dimensional lattice, {\it (ii)} $b=2$ which is
two dimensional but not a Bravais lattice, and {\it (iii)} $b=3$ as a
fractal-type lattice.


\hlattice


The hierarchical structure of the lattice allows us to calculate
$L(y)$ via a real space renormalization group approach, by decimating
spins on individual motifs\cite{hier,derrida}.  Let us define $Z_n=
2^N L_n$ and $f_n=(2b)^{-n}\ln Z_n$ for the $n$th generation.  Note
that $f_n(y)$ is related to $f(y)$ of Eq.  (\ref{eq:4}) by
$f=f_n(1)-f_n(y).$ $Z_n,$ and $f_n$ satisfy the following recursion
relations (see Supplemental Material \cite{suppl})
\begin{subequations}
\begin{eqnarray}
  \label{eq:7}
 Z_n(y)&=&\zeta(y_1) Z_{n-1}(y_1),~~\zeta(x)=2^b x^{1/2},\\
 f_n(y)&=& (2b)^{-1} f_{n-1}(y_1) +  
 (2b)^{-1} g(y_1),  \label{eq:b}
\end{eqnarray}
with $g(x)=\ln \zeta(x),$ and the RG flow equation
\begin{equation}
  \label{eq:8}
y_1={2^{-b}} ( y +{y}^{-1})^b.    
\end{equation}
The boundary conditions (BC) are encoded in $Z_1$ as,
\begin{eqnarray}
  \label{eq:2}
  Z_1=\left\{\begin{array}{ll}
              2(y^{1/2}+y^{-1/2}),& {\rm(Open~ BC)}\\
               y^{1/2}, & {\rm(Fixed ~BC)} (\uparrow\uparrow),
               \end{array}\right.
\end{eqnarray}
\end{subequations}
with $f_1=\ln Z_1$. 

Equation (\ref{eq:8}) has FPs at $y=1$ (infinite-temperature FP,
paramagnetic phase), $y=\infty$ (zero-temperature FP, ordered phase),
and a $b$-dependent unstable FP at $y=y_c$ (for $b>1$) representing
the critical point.  For any odd $b>1$, there are additional
``unphysical'' FPs at $y=-1, -y_c$ ($\pm\infty$ to be identified).
There is no $y_c$ for $b=1$, as there is no thermal phase transition
for the one-dimensional Ising model.  The zeros of $L_n(y)$ can be
determined from those of $L_{n-1}$ via Eqs.(\ref{eq:7}) and
(\ref{eq:8}), starting from the BC-dependent roots of $L_1(y)=0$.  In
the $n\to\infty$ limit, the zeros then belong to the set of points
that do not flow to infinity, thereby constituting the Julia set of
the transformation\cite{derrida}.  These sets, obtained by {\tiny
  MATHEMATICA}, are shown for $b=1,2,$ and $3$ in Figs.~\ref{fig:2}
and \ref{fig:22}.


\bzeros


The $y, y^{-1}$ symmetry in Eq.\,(\ref{eq:8}) ensures that if $y^*$ is a
FP, then $1/y^*$ flows to $y^*$.  Therefore, there are four special
points on the unit circle which flow to the nontrivial FPs, and are,
necessarily, members of the Julia set.  These four points on the unit
circle in Figs.~\ref{fig:2}(iii),\,2(iv), and Fig.~\ref{fig:22} 
are the four critical points
in time for $b>1$.  Incidentally, Eq.\,(\ref{eq:8}) also ensures that
any point on the unit circle, $y=e^{i\theta}$, under iteration, first
flows to the real axis to $\cos \theta$ and then remains real afterwards.
 Consequently, complex RG fixed points for $b>1$ are not
important.


\bjulia


DQPT has been studied for $b=1$ under periodic BC\cite{heylprl}.  The
surprising result we find here is that, unlike the thermal case,
boundary conditions may even suppress the bulk transition.  The
transfer matrix solution of the 1D Ising model describes the
partition function by the two eigenvalues $\Lambda_{\pm}=y^{1/2}\pm
y^{-1/2}$, with the larger one determining the $N\to\infty$
behaviour\cite{huang}.  For $y$ flowing to $y^*=+1~(y^*=-1)$, the
larger eigenvalue in magnitude is $\Lambda_{+}~(\Lambda_{-})$, so
that, with $y=e^{i\theta}$, the rate functions for the two regions
($f_{\pm}\sim \ln \Lambda_{\pm}/2$) are (see Supplemental material
\cite{suppl}) 
\begin{equation}
  \label{eq:5}
f_{+}(y)=-\frac{1}{2} \ln \cos^2 \frac{\theta}{2},
~~{\rm and}~~f_{-}(y)= -\frac{1}{2} \ln \sin^2 \frac{\theta}{2},
\end{equation}
respectively.  As characteristics of the high-temperature phases,
$f_{\pm}$ should be independent of dimensions, remaining valid for all
$b$.  Open BC yields only one zero at $y=-1$ [Fig.~\ref{fig:2}(i)],
and therefore no DQPT.  On the other hand, periodic and fixed BC give
zeros on the imaginary-$y$ axis [Fig.~\ref{fig:2}(ii)].  Two zeros
$y=\pm i$ on the unit circle, demarcating the RG flows of the points
on the unit circle to $y=\pm 1$, are the known transition
points\cite{heylprl,heyl2}.  The transitions are from a paramagnetic
(described by FP at $y=1$, and $f_+$) to another paramagnetic phase,
which we call para$^{\prime}$, described by FP $y=-1$, and rate
function $f_{-}$].
[Fig.~\ref{fig:4}(i)].

Now consider an open chain with the boundary term $H_{\rm
  B}=-h(\sigma_1^{\rm z}+\sigma_N^{\rm z})$.  For a finite chain,
there will be contributions from both the FPs $y=\pm 1$, so that for
an $N$-site chain (see Supplemental Meterial \cite{suppl})
\begin{equation}
  \label{eq:51}
  L(t,h)= (\cos Jt)^{N-1} \cos^2ht+(i \sin Jt)^{N-1}\sin^2ht. 
\end{equation}
DQPT with $f_{\pm}(t)$ is recovered in the $N\to\infty$ limit {\it
  only if} $h\neq 0$.  See Fig.~\ref{fig:4}(i).  For an open chain,
$L(t)\equiv L(t,{h=0})=(\cos Jt)^{N-1}.$ Hence, there is no transition
[Fig.\,\ref{fig:4}(ii)], consistent with one single zero
[Fig.\,\ref{fig:2}(i)].  There are four sectors of possible
configurations of the two boundary spins, viz., $(\pm,\pm)$.  Each of
these four sectors individually shows DQPT.  However, for the
zero-field open chain, requiring superposition of the four sectors,
there is a perfect cancellation of the $y=-1$ contributions.  Thus,
only $f_+$ survives [Fig.\,\ref{fig:4}(ii)].  When the subtle
cancellation of the four sectors is disturbed by the small boundary
fields, the transitions appear, as shown in Fig.~\ref{fig:4}(i).  We
see that boundary conditions (like open-chain) become relevant only at
the unphysical fixed point.

For any odd $b>1$, there are four critical points on the unit circle,
Figs.~\ref{fig:2}, and \ref{fig:22}.  These are
A$_1$, $\theta_A=2J\tau_1=\arccos {y_c}^{-1/b}$, and
A$_2$, $2J\tau_2=2\pi-\theta_A$ on the right half-plane, flowing to
$y_c>0$, and K$_1$, $2J\kappa_1=\pi-\theta_A=\arccos({-y_c})^{-1/b}$,
and K$_2$, $2J\kappa_2=\pi+\theta_A$, on the left half-plane, flowing
to the unphysical FP at $-y_c <0$, via $y=-1/y_c$.  These transition
points ($l\pi\pm\theta_A$, for any integer $l$) are determined
exactly.  In this particular case, the nature of the singularity
happens to be the same for all, as for the thermal case [a diverging
third derivative of $f$, Fig.\, \ref{fig:4}(iii)].  The flows of the
four arcs of the unit circle are shown in Fig.~\ref{fig:2}(vi).
K$_1$K$_2$, being characterized by FP $y=-1$, is expected to be
sensitive to any constraint on the boundary spins.  For, say, fixed
boundary spins, a sequence of phases occurs in time,
para-ferro-para$^{\prime}$-ferro-para, separated by the four
critical points.  The two para phases with FP $y=\pm 1$ have
ferromagnetic phases in between.  However, in the unbiased case, the
algebraic sum of the contributions of the four boundary sectors may
lead to cancellation as in the $b=1$ case.  A signature of the
cancellation in the K$_1$K$_2$ region is the failure of Eq.
(\ref{eq:b}) for $f$ as $y\to -1$ on renormalization.  This stability
problem is also seen in the one-dimensional case, Fig.~\ref{fig:4}(i)
{\it vis-\`a-vis} Fig.~\ref{fig:4}(ii) [the recursion relation, Eq.
(\ref{eq:b}), fails for $\pi/2<2Jt<3\pi/2$].  We, therefore,
conjecture that for the open BC case (free boundary spins) there is no
intermediate para$^{\prime}$ phase, but instead the whole arc
A$_1$K$_1$K$_2$A$_2$ represents the ferro phase---a major boundary
effect on bulk DQPT.


\onedchain

For even $b$, there are again four points on the unit circle
[Figs.~\ref{fig:2}{(iii)} and \ref{fig:22}{(i)}], A$_i, (i=1,4),$
which have identical angular relations as the four points for odd $b$,
except that here all flow to $y_c$ in two steps via $y=1/y_c$.  All
points in arcs A$_1$A$_4$ and A$_2$A$_3$ flow to $\infty$ implying an
ordered state, while the remaining two arcs, A$_1$A$_2$ and
A$_3$A$_4$, flow to $1$, the disordered phase.  Therefore, there is an
oscillation between ordered (broken-symmetry) phase and the standard
disordered phase with critical points at four different times.  The
nonanalytic features at the four critical times are the same as for
the temperature-driven critical point at $y_c$.  In essence, DQPT here
follows closely the thermal transition.

To show the generality of the boundary effect, let us consider the
three state Potts chain of $N$ sites (3QPC) involving $3\times 3$
matrices~\cite{ks17}. The interaction term is
\begin{equation}
  \label{eq:6}
H=-J\sum_{j} (\Omega^{\dagger}_j \Omega_{j+1} + {\rm H.c.}),   
\end{equation}
where $\Omega={\rm diag}(1,e^{i2\pi/3},e^{i4\pi/3})$. Analogous to the
transverse field of Eq. (\ref{eq:3}), the spin flipping term for Potts
spin is $H_{\Gamma}=-\Gamma \sum_j T_j$, where the elements of the
$3\times 3$
matrix $T$ are given by $T^{\alpha\beta}=1-\delta_{\alpha\beta},
(\alpha,\beta=1,2,3).$ $\Gamma$ can be used to prepare the chain in a
product state of equal-amplitude superpositions of the three states of
each spin.  The chain evolves in time with $H$ of Eq. (\ref{eq:6}), once
$\Gamma$ is switched off.  Two boundary conditions are considered
here, viz., periodic and open BCs.  These two differ by an interaction
term connecting the first and the $N$th sites. (See Supplemental
Material \cite{suppl}.)

In the Potts model the basic energy scale for a bond is the gap $3J$,
and so define $y=\exp(3\beta J)$.  The RG equation for $y$
is\cite{derrida,comm2}
\begin{equation}
  \label{eq:10}
y_1=R(y)\equiv (y^2+2)/(2y+1),
\end{equation}
whose fixed points are $y_p=1$, $y_u=-2$ (``unphysical''), and
$y=\pm\infty$.  The DQPT involves the transition between the two
stable phases described by $y_p$ and $y_u$, (analogous to
Fig.~\ref{fig:2}(iv)), with the critical times at the points of
intersection of the unit circle and the line of zeros of $L(y)=0$.
These intersectioons are A$_1$, $y_{A1}=e^{i2\pi/3}$, and A$_2$,
$y_{A2}=e^{i4\pi/3}$, which {\it flow into each other} under the RG
transformation.  In other words, A$_1$,i and A$_2$ are period-2 FPs of
RG, i.e., the {\it fixed points} of $R^{(2)}=R(R(y))$. By linearizing
$R^{(2)}$ around A$_1$ or A$_2$, the thermal eigenvalue is found to be
1, which leads to a kink in ${\rm Re}\,f$ at the transition points
\cite{ks17}.  The emergence of these novel ``unphysical'' fixed points
{\it distinguishes} the DQPT from thermal transitions in general, and,
in particular, the 3-state Potts chain from TFIM, though both show
similar nonanalyticity \cite{comm20}.

Now consider a free (i.e., open) chain. There is only one zero at
$y=-2$ outside the unit circle [compare with Fig.\ref{fig:2}(i)].
There cannot be any DQPT.   A direct computation of the rate function
$f(y)$ by the transfer matrix method\cite{ks17,huang} shows that
$f(y)\sim -\ln [(y+2)/3], (y=e^{i\theta}),$ and no DQPT, in agreement
with the zeros.  From the RG point of view, we see that the phase
described by the unphysical fixed point at $y=-2$ does not occur for
the open chain case, though it exists for the periodic case.  This
exact result provides yet another example of boundary conditions
affecting the bulk transition in the quantum case, when an unphysical
FP is involved.

Our results are summarized at the beginning, and, to that, we add the
following details.  The four transition points (critical times) for
dynamical quantum phase transitions are determined exactly for the
Ising model on hierarchical lattices of any $b>1$.  The ordered
(broken-symmetry) state appears as a phase only in higher dimensions
($b>1$).  For all odd $b$, the phase transitions involve one phase
characterized by a stable, but unphysical, fixed point.  There are no
such unphysical fixed points for even $b$, and, therefore, no
sensitivity to boundary conditions.  We anticipate that our results
would lead to explorations of other unphysical fixed points in various
quantum systems to look for new phases, criticality, and boundary
effects, not found in thermal phase transitions\cite{comm3}.

\begin{acknowledgments}
{A. K. thanks Sk. Sazim for discussions on MATLAB. S. M. B. acknowledges support
from the J. C. Bose Fellowship Grant (DST). }
\end{acknowledgments}

{\it Note added.}---Recently, Ref. \cite{comm3} appeared, which dis-
cussed the role of complex fixed points for Potts chains.

\vfill
\newpage

\onecolumngrid

\begin{center}
\textbf{\large Supplemental  Material on}

\textbf{\large {``Boundaries and Unphysical Fixed Points in Dynamical Quantum
  Phase Transitions"}} 

\textbf{Amina Khatun$^1$ and Somendra M. Bhattacharjee$^2$}

$^1$ Institute of Physics, Bhubaneswar 751005, India\\
$^2$ Department of Physics, Ashoka University, Sonepat, Haryana - 131029, India
 
\end{center}
\setcounter{equation}{0}
\setcounter{figure}{0}
\setcounter{table}{0}
\setcounter{page}{1}
\setcounter{section}{0}
\renewcommand{\theequation}{S\arabic{equation}}
\renewcommand{\thefigure}{S\arabic{figure}}
\renewcommand{\thesection}{S\arabic{section}}
\renewcommand{\bibnumfmt}[1]{[S#1]}
\renewcommand{\citenumfont}[1]{S#1}

\vspace{12pt}
\maketitle
\twocolumngrid
\section{I.  On hierarchical latices}

For a hierarchical lattice (diamond type) of $b$ branches, the
dimension\cite{Ssmb} is given by 
\begin{equation}
  \label{eq:10s}
d= \frac{\ln 2b}{\ln 2},  
\end{equation}
so that $d=1,~2$ for $b=1,~2$, while $d=\ln 6/\ln 2=2.58...$ for
$b=3$.  For the $n$th generation, the number of bonds is $B_n=(2b)^n$
and the number of sites is
\begin{equation}
  \label{eq:11}
N_{n}= b B_{n-1}+N_{n-1}=2+ b \frac{(2b)^n-1}{2b-1}.  
\end{equation}
In the $n\to\infty$ limit, $\frac{N_n}{B_n} \to \frac{b}{2b-1}.$

\section{II.  Partition function}

The partition function of the traditional nearest neighbour Ising model 
is given by\cite{Shuang}
\begin{equation}
  \label{eq:s8}
Z~=~\sum_{C\in 2^N {\rm states}} e^{-\beta E_C},  
\end{equation}
with $$E_C=-J\sum_{\langle jk\rangle} s_j s_k,$$
 as the energy of
configuration C.  Here, $s_j=\pm 1$, $\beta=1/k_BT$, $T$ being the
temperature and $k_B$ the Boltzmann constant.  The free energy is
given by $-k_BT \ln Z$.

We take $y=e^{2\beta J}$ as the variable because $2J$ is the
energy gap for a single bond. 

The existence of the thermodynamic limit ($N\to\infty$) for the Ising
model ensures the large deviation form $L~=~\exp(-Nf(y))$, with $f$ as
the analog of the free energy extended to the complex plane.

\section{RG relations for the Ising model}

\begin{figure}
\includegraphics[clip]{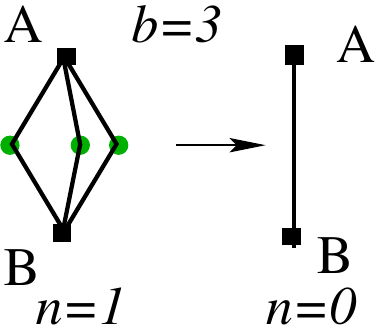}
\caption{Decimation of a motif of $b$ branches by summing over the
  internal spins (green disks), keeping the spins at A and B fixed.}\label{fig:s1}
\end{figure}

We derive the RG relations for the Ising model on a hierarchical
lattice of $b$ branches.  The decimation of a motif to a bond is shown in
Fig. \ref{fig:s1}

Denoting the partition function of a bond of two spins $s_1,s_2=\pm 1$
by $Z_{s_1s_2}(y)$, the summation over the internal spins (green
disks) yields the relation between the partition function $Z_1$ of the
larger cell (generation $n=1$) as a function of $y$ and the partition
function of a bond ($n=0$) as a function of the renormalized parameter $y_1$.
Explicitly,
\begin{subequations}
\begin{eqnarray}
  \label{eq:s2}
  Z_1|_{++}&\equiv&\left[(Z_{++}(y))Z_{++}(y)+
    Z_{+-}(y)Z_{-+}(y)\right]^{b}\nonumber\\
     &=&\zeta(y_1)\;  Z_{++}(y_1),\\
 Z_1|_{+-}&\equiv&\left[Z_{+-}(y)  Z_{--}(y)+Z_{++}(y)
   Z_{+-}(y)\right]^{b}\nonumber\\
        &=&\zeta(y_1)\;  Z_{+-}(y_1),
\end{eqnarray}
\end{subequations}
exploiting the fact that the $b$ branches are independent.  Only two
relations are sufficient, thanks to the symmetry that $Z_{++}(y)=
Z_{--}(y),$ and $Z_{+-}(y)= Z_{-+}(y).$ Noting that
$Z_{++}(y)=y^{1/2},$ and $Z_{+-}(y)=y^{-1/2}$, one gets
\begin{subequations}
\begin{eqnarray}
  \label{eq:s3}
  y_1&=&2^{-b} (y + y^{-1})^b,\\
  \zeta(y)&=&2^b y^{1/2}.
\end{eqnarray}
\end{subequations}

\section{III. Derivation of $f_{\pm}$ and Zeros of the one-dimensional Ising  chain}
The two high temperature fixed points of Eq. \ref{eq:s3},
$y^*=\pm 1$ characterize the two possible phases of the one dimensional
Ising model.
The Ising partition function can be determined by a transfer matrix
approach \cite{Shuang}.

If ${\sf T}$ is the $2\times 2$ transfer matrix, then the
partition functions for a chain of $N$ sites under different BCs are given by
\begin{equation}
  \label{eq:s2ising}
  Z_N=\left\{ \begin{array}{ll}
               {\rm Tr}~ {\sf T}^N,& {\rm(periodic\  BC),}\\
                \sum_{j,\;k=1,\;2}~ [{\sf T}^{N-1}]_{jk},& {\rm(free\  BC),}\\
               (e^{zh}\, e^{-zh}) {\sf T}^{N-1}
               \left({{\rm e}^{zh}}\atop{e^{-zh}}\right ), &{\rm(with\
                 boundary\  fields),}
               \end{array}\right.
\end{equation}
For periodic BC
\begin{equation}
  \label{eq:s41}
Z_N=\Lambda_{+}^N+\Lambda_{-}^N,  
\end{equation}
 where $\Lambda_{\pm}=y^{1/2}\pm y^{-1/2}$ are the two eigenvalues of
 ${\sf T}$.   Eq. 6 of the text  follows from Eq. (\ref{eq:s2ising})

 For large $N$, and $y=e^{i\theta}$
\begin{equation}
  \label{eq:s5}
Z_N =\left\{\begin{array}{ll}
              \Lambda_{+}^N,& {\rm for}\  \theta \ {\rm near}\  0,\ |\Lambda_+|>|\Lambda_{-}|\\
              \Lambda_{-}^N,& {\rm for}\  \theta \ {\rm near}\  \pi,\ |\Lambda_-|>|\Lambda_{+}|
              \end{array}\right.
\end{equation}
The rate functions 
$$f_{\pm}(\theta)\equiv - {\rm Re} \lim_{N\to\infty} \ln (Z_N/2^N)=-
{\rm Re}\ln(\Lambda_{\pm}/2),$$
 are then given by 
\begin{equation}
  \label{eq:s4}
  f_+=-\frac{1}{2}\ln \cos^2 \frac{\theta}{2},~~{\rm and}~~
  f_-=-\frac{1}{2}\ln \sin^2 \frac{\theta}{2},
\end{equation}
as quoted in Eq. (5) of the text.

{\bf Zeros of the one-dimensional Ising  chain}

The partition function for an $N$-site Ising  chain with periodic 
boundary condition is given in Eq. (\ref{eq:s41}).  The zeros of
$L_N(y)=Z_N/2^N$ are then given by
$$y=i~ {\rm cot} \frac{(2n+1)\pi}{N}, n=-N,...,N-1,$$
 which lie along the imaginary axis.
This is shown in Fig 2(b) in the text.

For open BC, the partition function is given by $Z_N=\Lambda_{+}^N$.
There is therefore only one zero at $\Lambda=0$.  Therefore, $L_N(y)$
has only  one zero at $y=-1$, as shown in Fig 2(a) in the text.

\section{V.  Zeros of the one-dimensional  Potts  chain}

For the Potts model $y=\exp(3\beta J)$.  
The partition functions for the Potts chain are given by
\begin{equation}
  \label{eq:s82}
  Z_N= \left\{ \begin{array}{ll}
               \Lambda_1^N+2\Lambda_2^N, &\quad(\rm periodic~BC),\\
               \Lambda_1^N, &\quad(\rm Open~ BC),
               \end{array}\right. 
\end{equation}
where $\Lambda_1\propto (y+2),$ and  $\Lambda_2\propto (y-1)$ are the
two eigen values of the transfer matrix given in Ref. \cite{sks17}.  The
zeros for the periodic BC case are given by $\Lambda_1/\Lambda_2=
e^{i (2n+1)\pi/N}$, so that 
\begin{equation}
  \label{eq:s81}
y=-\frac{1}{2} + i \frac{3}{2} {\rm cot} \frac{\pi (2n+1)}{N}, \quad {\rm(PBC)}. 
\end{equation}
For an open chain (free BC), it follows from Eq. (\ref{eq:s82}) that
there is only one zero at $\Lambda_1=0$, i.e., at $y=-2$.

\section{V.  On $L_n$ and  $f_n$}

 Under the decimation transformation of Fig. \ref{fig:s1} and Eq.
  (\ref{eq:s2}), the partition function for generation $n$ with
  parameter $y$, $Z_n(y)$, can be related to that of the $(n-1)$th
  generation with parameter $y_1$.  The recursion relations for $Z_n$,
  and $f_n=(2b)^{-n} \ln Z_n$  can be written as\cite{Sderrida}
\begin{eqnarray}
  \label{eq:s6}
Z_n(y)&=&\zeta(y_1) Z_{n-1}(y_1),\\
~{\rm and}~ f_n(y)&=& \frac{1}{2b} f_{n-1}(y_1) +   \frac{1}{2b} g(y_1),    
\end{eqnarray}
where $g(x)=\ln \zeta(x)=2^{-1} \ln(4^b x)$, and $y_1$ is given by 
Eq.~(\ref{eq:s3}).
Note that  $f_n$ is  defined without the normalization factor $2^N$
(Eq.\,(4a) in text). With successive 
transformation $y\to y_1\to...\to y_n$ following the RG flow equation, the
Loschmidt amplitude is given by a rapidly convergent sum for large $n$
as
\begin{equation}
  \label{eq:s7}
  f_{n+1}(y)=\sum_{j=1}^{n} \frac{1}{(2b)^j} g(y_j) + \frac{1}{(2b)^{n}}
  f_1(y_{n}),    
\end{equation}
provided the functions remain well-defined at the transformed
arguments.  

 The fixed points of Eq.(\ref{eq:s3}) satisfy the equation 
\begin{equation}
  \label{eq:12}
y^2 -2 y^{(b+1)/b} +1=0.  
\end{equation}
The nontrivial fixed points are $y_c = 3.38298.... $ for $b=2$, and 
$y_c = 2.05817...$ for b = 3.

 In the limit $n\to\infty$,  one gets 
\begin{equation}
  \label{eq:13}
f_{\infty}(1)=\frac{b}{2b-1}\, \ln 2,  
\end{equation}
which is the infinite temperature entropy per bond Eq. (\ref{eq:11})). 
The rate function $f$
as defined in Eq.\,~(4a) in the text, is given by  $f(y)\equiv f_n(1)-f_n(y)$,
where $f_n(1)$ 
takes care of the normalization in $L$. 
Fig.~4 in the text shows the plots of $f(y)$.

 For points $A_i$'s and $K_i$'s that flow to $\pm y_c$ in two steps, we
have the same value of $f=f_c$ for all of them with
\begin{equation}
  \label{eq:1}
  f_c= \frac{1}{4b}\ln \frac{4^b}{y_c} + \frac{1}{4b(2b-1)} \ln (4^b
  y_c) -\frac{b}{2b-1}\, \ln 2,
\end{equation}
which evaluates to 
$$f_c\big|_{b=2}=0.10156312,\quad {\rm and}\quad f_c\big|_{b=3}=0.0482183. $$
The intersection of the  $f=f_c$ line with the $f$-vs-$t$ curve gives the
critical points $A_i$'s and $K_i$'s.

For $b=3$, the transition points are
\begin{eqnarray*}
{\rm A}_1&:& 2J\tau_1=0.666239...,\\
{\rm A}_2&:& 2J\tau_2=2\pi-0.666239=5.616946,\\
{\rm K}_1&:& 2J\kappa_1=\pi-0.666239=2.47535,\\
{\rm and~ K}_2&:& 2J\kappa_2=3.807832.
\end{eqnarray*}

For $b=2$, the transition points are
\begin{eqnarray*}
{\rm A}_1&:& 2J\tau_1=0.99597...,\\
{\rm  A}_2&:& 2 J\tau_2=5.28722,\\
{\rm A}_3&:& 2 J\tau_3=4.13756,\\
{\rm and~~A}_4&:& 2 J\tau_4=2.14562.
\end{eqnarray*}


\begin{thebibliography}{99}
\bibitem{comm1} The periodic nature, when it happens, is one of the
  dramatic aspects of the transition. It  should be
  mentioned that DQPT is not necessarily periodic in every 
  situation.
\bibitem{comm11}  DQPT is different from dynamical
transitions or dynamics of usual phase transitions, which are all
bereft of any special time \cite{cardy}.
\bibitem{spectrum}
For a lattice of $N$ sites, the complete
 spectrum, $H|\phi_C\rangle=E_C|\phi_C\rangle$, is given by
$   |\phi_C\rangle=\bigotimes_{j=1,N} |\bullet\rangle_j,$  and
$E_C=-J\sum_{\langle jk\rangle} s_j s_k,$ where $|\bullet\rangle_j$ is
 either $|\!\uparrow\rangle$ or $|\!\downarrow\rangle$ eigenstate of
 $\sigma^z_j$ with $s_j=\pm 1$ in configuration $C$ (there are $2^N$
 states).
\bibitem{commx} The initial state is $\psizk~=~\bigotimes_j
  |\!\to\rangle_j$, where
  $|\!\to\rangle$
  is an eigenstate of $\sigma^{\rm x}$.  For TFIM, the paramagnetic
  and the ferromagnetic states are defined with respect to the net
  orientation in the z-direction. The state with all spins in the
  x-direction, by this definition, is a para state.
\bibitem{comm2} The RG equation can be obtained by comparing $M^2$ with
  $M$ where $M$ is the transfer matrix of Ref.\,\cite{ks17}.
\bibitem{comm20} More details on the Potts model for different $b$
  will be reported elsewhere.
\end{thebibliography}
\end{document}